\documentclass[12pt]{article}

\usepackage[a4paper,margin=1in]{geometry}
\usepackage[T1]{fontenc}
\usepackage[utf8]{inputenc}
\usepackage{lmodern}
\usepackage{microtype}
\usepackage{tabularx}
\newcolumntype{L}{>{\raggedright\arraybackslash}X}
\usepackage{graphicx}
\usepackage{amsmath,amssymb,amsfonts,mathtools,bm}
\usepackage{amsthm}
\usepackage{siunitx}
\usepackage{enumitem}
\usepackage{upgreek}
\usepackage{mathrsfs}
\usepackage{booktabs}
\usepackage{array} 
\usepackage{hyperref}
\usepackage[nameinlink,capitalise]{cleveref}

\hypersetup{
  colorlinks=true,
  linkcolor=blue,
  citecolor=blue,
  urlcolor=blue
}

\newtheorem{proposition}{Proposition}[section]
\newtheorem{remark}[proposition]{Remark}

\title{Evidence for Bures--Wasserstein Boundary Dynamics\\
in the Living Human Brain}
\author{Christian Kerskens}
\date{\today}

\begin{document}
\maketitle

\begin{abstract}
When substrate-constrained covariance flow on the Bures--Wasserstein
manifold reaches the Williamson boundary, single-mode compression
saturates and further admissible covariance evolution is forced into the
cross-mode complement. This paper derives how that substrate boundary
transition becomes experimentally visible in an embedded spin probe in
the living human brain.

We formulate a boundary-conditioned transfer theorem: when the
substrate enters the deep boundary regime in a coupled mode, the
boundary-selected cross-mode continuation of substrate covariance flow
enters the reduced spin dynamics as a nonzero inter-spin correlation
block. The spin probe does not inherit the substrate boundary as a
state; it detects the boundary indirectly through the transferred
cross-mode sector of the reduced dynamics. To leading order, this
transfer is selective: it acts through an additive cross-diffusion
channel while leaving conventional single-mode NMR observables such as
\(T_1\), \(T_2\), linewidths, and the ordinary single-quantum response
dominated by the thermal background.

Projecting the induced spin cross-mode structure into the two-spin
algebra, we argue that the experimentally relevant dominant recipient is
the double-quantum SU(1,1) pair sector rather than the compact
zero-quantum SU(2) exchange sector. We then derive the
coherence-transfer pathway through which this double-quantum pair
coherence is converted into a detectable signal by the
\(45^\circ\)--gradient--\(45^\circ\) readout block.

Previously published brain magnetic-resonance data exhibit signatures
consistent with the predicted phenomenology of this mechanism:
balanced preparation, immediate onset, refocusing at the pair frequency
\(\Omega_+\), even-echo structure, and incompatibility with compact
SU(2) exchange. The detected signal is therefore interpreted first as a
witness of entry into the deep Bures--Wasserstein boundary regime, and
second as a macroscopic witness of SU(1,1)-type pair-sector
multiple-quantum coherence and squeezing. Finally, we formulate the
entanglement-theoretic framework required to elevate this signal to a
many-body entanglement witness, and identify the quantitative
conditions required for that future closure.
\end{abstract}

%======================================================================
\section{Introduction}
\label{sec:intro}
%======================================================================

This paper is the third in a programme connecting
Bures--Wasserstein covariance geometry to experimentally observable
signatures in the living human brain. The first paper~\cite{Paper1} develops
the Bures--Wasserstein geometry of Gaussian covariance matrices and
establishes the constructive noise bound that governs cross-mode
structure at the Williamson floor. The companion
paper~\cite{MetricPaper} applies that geometry to neural substrate
dynamics, showing that substrate-constrained covariance flow can drive a
neural mode to the Williamson boundary, where single-mode compression is
exhausted and further admissible covariance evolution is forced into the
cross-mode complement.

The present paper addresses the measurement problem left open by those
results. Clinical magnetic resonance does not observe the substrate
covariance directly. It observes a spin probe embedded in the
substrate. The question is therefore: if the neural substrate enters
the deep Bures--Wasserstein boundary regime predicted by the companion
paper, what should an embedded spin probe detect?

The central claim of the paper is that the spin probe does not detect
the substrate boundary by copying the substrate state. Nor does the
reduced spin system need to reach the same Williamson-boundary point as
the substrate. What is transferred is more specific: once the substrate
reaches the Williamson boundary, the admissible continuation of its
covariance flow becomes cross-mode in the saturating sector, and this
boundary-conditioned cross-mode structure enters the reduced spin
dynamics as a nonzero inter-spin correlation block. The spin probe
therefore registers the substrate boundary indirectly, through the
transferred cross-mode sector of its reduced generator.

This transferred cross-mode structure must then be identified
algebraically. The two-spin operator algebra admits both a compact
zero-quantum SU(2) exchange sector and a non-compact double-quantum
SU(1,1) pair sector. The projection formulas derived below show how the
transferred spin cross-mode block decomposes into these candidate
recipients, while the experimental refocusing at the pair frequency
\(\Omega_+\), rather than the exchange frequency \(\Omega_-\), identifies
the dominant experimentally relevant recipient as the double-quantum
SU(1,1) pair sector.

A second problem then arises. If the dominant recipient is the
double-quantum SU(1,1) pair sector, the relevant coherence is not
directly preserved by the gradient-filtered readout used in the original
experiment~\cite{KerskensPerez2022}. The signal must therefore be
understood as a readout-converted output of the pair sector rather than
as a direct observation of it. We derive the corresponding
coherence-transfer pathway explicitly and show that the
\(45^\circ\)--gradient--\(45^\circ\) block converts a portion of the
double-quantum pair coherence into a gradient-surviving intermediate and
then into detectable single-quantum magnetization.

This logic yields a continuous chain from geometry to experiment:
\[
\begin{aligned}
  &\text{substrate BW boundary} 
  \to \text{boundary-conditioned cross-mode continuation} \\
  &\quad \to \text{transferred spin cross-mode structure} 
  \to \text{DQ/SU(1,1) pair-sector projection} \\
  &\quad \to \text{readout-converted magnetic-resonance signal}.
\end{aligned}
\]

Throughout, the primary claim is covariance-geometric rather than
microscopic: the spin probe is not taken to reveal the detailed material
constitution of the substrate, but to register the fluctuation
statistics imposed when the substrate enters a boundary-limited
informational regime.

The distinctive claim of the paper is that the observed signal is best
understood as the spin-probe signature of this full chain, not as an
isolated pulse-sequence artifact and not as a conventional compact
exchange phenomenon.

The reported magnetic-resonance data~\cite{KerskensPerez2022} exhibit
precisely the features predicted by this mechanism: balanced
preparation, immediate onset, refocusing at the pair frequency
\(\Omega_+\), even-echo structure, approximate linear scaling with
\(M_0\), and incompatibility with compact SU(2) exchange. Taken
together, these features support the interpretation that the detected
signal is an observable signature of boundary-conditioned transfer of
cross-mode structure from the neural substrate into the spin probe.

Accordingly, the signal is interpreted through a three-level hierarchy.
At the weakest level, it is a metric-regime witness: it indicates entry
into the deep Bures--Wasserstein boundary regime in which single-mode
compression is exhausted and cross-mode continuation is forced. At the
next level, it is an SU(1,1)-pair-sector multiple-quantum-coherence and
squeezing witness: it indicates nonzero collective double-quantum pair
coherence in the reduced spin system. At the strongest level, and only
under additional quantitative conditions, it may be promoted to a
many-body entanglement witness within the macroscopic MQC framework.
Because the room-temperature bulk-NMR setting is identity-dominated, a
strictly bipartite reduced-spin witness is thermodynamically
obstructed; the appropriate entanglement-theoretic setting is therefore
collective rather than isolated-pair.

At present, the quantitatively established claims of the paper are the
metric-regime and pair-sector MQC/squeezing levels of this hierarchy,
while the many-body entanglement upgrade remains a formal framework
awaiting numerical closure.

The structure of the paper is as follows. \Cref{sec:substrate}
summarises the companion paper's result that substrate-constrained
covariance flow reaches the Williamson boundary and is then forced into
the cross-mode complement. \Cref{sec:bridge} derives the
boundary-conditioned transfer theorem connecting substrate geometry to
reduced spin observables. \Cref{sec:algebra} classifies the relevant
two-spin subalgebras and identifies the non-compact SU(1,1) pair sector
as the dominant algebraic recipient of the transferred cross-mode
structure. \Cref{sec:pathway} derives the coherence-transfer pathway
through which the double-quantum pair sector becomes experimentally
visible. \Cref{sec:preptrain} interprets the repetitive preparation
train as a discrete flow on the spin Bures--Wasserstein manifold.
\Cref{sec:evidence} compares the resulting predictions with the reported
experimental features. \Cref{sec:witness} establishes the three-level
witness hierarchy. \Cref{sec:bipartite} distinguishes collective and
two-subsystem realizations of the SU(1,1) generators. Finally,
\Cref{sec:discussion} discusses the implications, limitations, and
falsifiability of the proposed mechanism.

%======================================================================
\section{Substrate Boundary Dynamics: Summary of the Companion Result}
\label{sec:substrate}
%======================================================================

This section summarises the result of the companion
paper~\cite{MetricPaper} that serves as the starting point for the
present analysis. The full derivation belongs to that paper; here we
state only the ingredients needed to connect the substrate geometry to
the spin-probe signal.

\subsection{Substrate covariance on the Bures--Wasserstein manifold}

The companion paper models the relevant neural substrate through a
covariance matrix
\[
  \Sigma_{\mathrm{sub}} \in \mathrm{Sym}^+(2n),
\]
representing the second-order statistical structure of the embodied
substrate dynamics on the Bures--Wasserstein manifold. At the present
level of description, the individual microscopic carriers of this
covariance need not be specified; what matters is that the substrate
admits a physically embodied covariance geometry whose admissible flow is
constrained by the Williamson boundary.

The corresponding Williamson decomposition is
\begin{equation}
  \Sigma_{\mathrm{sub}}
  =
  S^{T}\,
  \mathrm{diag}(\nu_1,\nu_1,\nu_2,\nu_2,\dots)\,
  S,
  \qquad
  S\in\mathrm{Sp}(2n,\mathbb{R}),
  \label{eq:substrate:williamson}
\end{equation}
where the symplectic eigenvalues \(\nu_k\geq \hbar/2\) encode the
irreducible uncertainty content of the normal modes. The value
\(\nu_k=\hbar/2\) is the Williamson floor, i.e.\ the minimum-uncertainty
bound for the corresponding physical mode.

\subsection{Substrate-constrained gradient flow}

The companion paper shows that the substrate covariance evolves under a
substrate-constrained gradient flow on \(\mathcal{BW}(2n)\), subject to
physiological constraints such as metabolic cost, anatomical embedding,
and finite transport capacity. The central result is that one or more
symplectic eigenvalues can be driven toward the Williamson floor:
\begin{equation}
  \nu_k^{\mathrm{sub}}(t)
  \longrightarrow
  \frac{\hbar}{2}
  \qquad
  \text{for the fastest-compressing mode }k.
  \label{eq:substrate:floor}
\end{equation}

This is the key boundary result imported from the companion paper. It
says that the substrate flow can reach a point at which single-mode
compression in the leading mode is exhausted.

\subsection{What changes at the boundary}

The important consequence is not merely that a symplectic eigenvalue has
become small, but that the admissible continuation of the flow changes
qualitatively at the boundary. Once a mode reaches the Williamson
boundary, further inward motion in that single-mode direction is no
longer admissible. If covariance reduction continues, it must be
redirected into the cross-mode complement.

This is the point at which the constructive noise bound from the first 
paper~\cite{Paper1} becomes decisive. In the notation of that paper,
continued admissible covariance evolution at the boundary requires
nonzero cross-mode structure, bounded below by
\begin{equation}
  \|G_{AB}\|
  \geq
  2\,\nu_{\min}\,|g|.
  \label{eq:substrate:cnb}
\end{equation}
Thus the companion paper establishes the following geometric fact:
\begin{quote}
\emph{At the Williamson boundary, further admissible substrate
covariance evolution is forced out of the saturating single-mode sector
and into the cross-mode sector.}
\end{quote}

\subsection{What the present paper takes from the companion result}

The present paper does \emph{not} assume that the spin probe simply
copies the substrate state. Nor does it assume that the reduced spin
covariance itself reaches the same boundary point. What is imported from
the companion paper is more precise: the substrate boundary selects a
specific \emph{cross-mode continuation} of covariance flow. That
boundary-conditioned cross-mode structure is the object whose transfer
to the spin probe will be analysed in \Cref{sec:bridge}.

Accordingly, the logical role of the companion paper is the following:
\begin{enumerate}[label=(\roman*)]
  \item it identifies the substrate as a physical system whose covariance
  evolves on the Bures--Wasserstein manifold;
  \item it shows that the substrate-constrained flow can reach the
  Williamson boundary in a mode;
  \item it proves that, at that boundary, continued admissible evolution
  must become cross-mode.
\end{enumerate}

The question of the present paper is therefore not whether the substrate
reaches the boundary --- that is the companion result --- but how the
boundary-conditioned cross-mode sector of the substrate becomes visible
in the reduced dynamics of an embedded spin probe.

\paragraph{Ontological status of the boundary regime.}
It is important to clarify the ontological status of the
boundary-conditioned transfer described in this framework. The
Bures--Wasserstein boundary is not interpreted here as a literal
macroscopic freezing or rigid deformation of neural tissue. Rather, it
represents an informational and covariance-geometric limit of the
substrate dynamics: the point at which independent single-mode
uncertainty reduction is no longer admissible and further covariance
evolution must be redirected into the cross-mode sector.

Because this informational dynamics is physically embodied, the
constraint cannot remain purely abstract. It must appear in the
statistics of the underlying biological fluctuations. In the present
framework, the boundary therefore manifests physically not as a bulk
mechanical singularity, but as a reorganization of the shared
fluctuation environment seen by the spin probe. The probe does not
measure the informational boundary directly as a classical thermodynamic
variable; rather, it registers the altered cross-correlated noise floor
\(D_{AB}\) induced when the substrate enters this boundary-limited
regime.

In this sense, the NMR signal is interpreted as a physical witness of a
covariance-geometric informational state, accessed indirectly through
its consequences for the fluctuation statistics of the embodied
substrate.

\subsection{Transition to the spin-probe problem}

This is the precise point at which the measurement problem begins.
Clinical magnetic resonance does not observe the substrate covariance
directly. It observes a spin probe embedded in the substrate. The issue
is therefore how a boundary-conditioned geometric change in the
substrate covariance flow enters the reduced spin dynamics and becomes
an experimentally detectable signal.

The next section derives exactly that step. Its central claim is that
the spin probe detects the substrate boundary not by inheriting the
boundary state, but by inheriting the \emph{boundary-conditioned
cross-mode continuation} of the substrate flow.

%======================================================================
\section{Boundary-Conditioned Transfer: From Substrate Geometry to Spin Correlations}
\label{sec:bridge}
%======================================================================

The preceding section established that the substrate-constrained
gradient flow can drive \(\Sigma_{\mathrm{sub}}\) to the Williamson
boundary in one mode, and that further admissible continuation is then
forced into the cross-mode complement. We now derive the consequence for
a spin probe embedded in that substrate. The central claim is that what
is transferred is not a copied boundary state but the
boundary-conditioned cross-mode continuation of the substrate flow.

\begin{remark}[Meaning of ``boundary-conditioned'' ]
\label{rem:boundaryconditioned}
We call a contribution to the reduced spin dynamics
\emph{boundary-conditioned} if it is driven by the cross-mode projection
of the substrate covariance-flow tangent vector,
\[
\Pi_{\mathrm{cross}}\dot{\Sigma}_{\mathrm{sub}},
\]
rather than by the purely marginal part of the substrate flow. In the
purely marginal interior regime this contribution vanishes, while in the
deep boundary regime --- where further marginal compression becomes
kinematically obstructed --- it becomes nonzero and enters the reduced
spin dynamics through the cross-mode sector.
\end{remark}

\subsection{Reduced open-system setting}

We model the spin probe as an open quantum system whose environment is
the neural substrate. In the Gaussian second-moment regime, the reduced
spin state is characterised by a covariance matrix
\[
  \Sigma_{\mathrm{spin}}\in\mathrm{Sym}^+(2m),
\]
whose dynamics are generated by a drift matrix \(A\) and a diffusion
matrix \(D\). At the level of second moments, one may write a continuous
Lyapunov equation of the form
\begin{equation}
  A\,\Sigma_{\mathrm{spin}}
  + \Sigma_{\mathrm{spin}}A^T
  + D
  = 0,
  \label{eq:lyap}
\end{equation}
for the stationary covariance, or the corresponding time-dependent
equation for the driven case.

The key structural point is that the diffusion matrix itself has a block
decomposition,
\begin{equation}
  D
  =
  \begin{pmatrix}
    D_A & D_{AB} \\
    D_{AB}^T & D_B
  \end{pmatrix},
  \label{eq:D:block}
\end{equation}
where \(D_A\) and \(D_B\) govern the marginal single-spin noise sectors,
while \(D_{AB}\) governs the correlated noise entering the inter-spin
sector. Conventional NMR observables such as \(T_1\), \(T_2\), and the
ordinary linewidths are controlled primarily by the marginal blocks
\(D_A\) and \(D_B\). By contrast, nontrivial pair-sector structure is
driven by the cross block \(D_{AB}\).

This observation already indicates what the bridge mechanism must and
must not do. If the substrate boundary were communicated to the spin
probe by driving the marginal diffusion blocks \(D_A\) and \(D_B\) to
their own quantum floor, one would expect dramatic changes in ordinary
relaxation observables and in the conventional single-quantum response.
That is neither the distinctive experimental signature of interest nor
the mechanism developed here. The present proposal is instead that the
substrate boundary appears in the spin probe primarily through
activation of the cross-mode sector.

\subsection{Shared-environment coupling and the origin of spin cross-diffusion}
\label{sec:bridge:coupling}

The role of the following open-system construction is not to replace the
covariance-geometric description with a microscopic tissue model, but to
show how a covariance-geometric constraint can become visible in a
physically embodied probe. The shared-environment reduction should
therefore be read as an observability map from substrate covariance
geometry to reduced spin fluctuation statistics.

The boundary-conditioned transfer mechanism requires a physical map from
the substrate covariance flow to the reduced spin diffusion tensor. This
map cannot simply be postulated by defining the boundary-conditioned part
of the spin generator to be whatever responds to the substrate
cross-mode sector. It must be derived from the fact that the two spin
subsystems are embedded in, and coupled to, the same underlying neural
substrate.

Let the two spin subsystems \(A\) and \(B\) be linearly coupled to a
common Gaussian substrate. At the level of system--bath coupling, write
the interaction schematically as
\[
H_{\mathrm{int}}
=
\sum_{\alpha}
X_A^{(\alpha)} \otimes F_A^{(\alpha)}
+
\sum_{\beta}
X_B^{(\beta)} \otimes F_B^{(\beta)},
\]
where the \(X\)'s act on the spin probe and the \(F\)'s are substrate
force operators. In the standard weak-coupling Markovian treatment of
Gaussian open systems, the reduced diffusion tensor is determined by the
symmetrized bath correlation kernel,
\begin{equation}
  D_{ij}
  =
  \int_0^\infty d\tau\;
  \sum_{\mu,\nu}
  C_{i\mu}\,
  \Gamma_{\mu\nu}(\tau)\,
  C_{j\nu},
  \label{eq:bridge:Dkernel}
\end{equation}
where \(C\) is the coupling matrix and
\[
\Gamma_{\mu\nu}(\tau)
=
\tfrac12
\bigl\langle
\{F_\mu(\tau),F_\nu(0)\}
\bigr\rangle
\]
is the symmetrized substrate correlation kernel
\cite{BreuerPetruccione2002,WisemanMilburn2010}.

For the two-subsystem decomposition, the diffusion tensor takes the block
form
\[
D
=
\begin{pmatrix}
D_A & D_{AB}\\
D_{AB}^T & D_B
\end{pmatrix}.
\]
The marginal blocks \(D_A,D_B\) are determined by the autocorrelation
kernels of the substrate forces acting separately on \(A\) and \(B\),
whereas the cross block \(D_{AB}\) is determined by the
\emph{cross-correlation} kernel shared between the two subsystems.

In the local linear-Gaussian approximation, this relation reduces
schematically to a covariance projection of the form
\begin{equation}
  D^{(\mathrm{BW})}
  \sim
  C\,N_{\mathrm{sub}}\,C^T,
  \label{eq:bridge:Dlocal}
\end{equation}
where \(N_{\mathrm{sub}}\) is the effective equal-time noise covariance
of the substrate sector seen by the spins. In particular, the
boundary-conditioned cross block obeys
\begin{equation}
  D_{AB}^{(\mathrm{BW})}
  \sim
  C_A\,G_{AB}^{\mathrm{sub}}\,C_B^T,
  \label{eq:bridge:DABmap}
\end{equation}
up to the effective kernel reduction appropriate to the Markovian
approximation.

Equation~\eqref{eq:bridge:DABmap} provides the missing physical link:
the spin cross-diffusion block is not defined ad hoc by the substrate
cross-mode tangent sector; it is generated by the cross-correlated
substrate fluctuations seen jointly by the two spins.

\paragraph{Approximation class.}
We work in a weak-coupling shared-environment reduction of the spin
probe, with the conventional marginal bath treated in the Born--Markov
regime. The bath correlation time is assumed to be short compared with
the effective spin-relaxation timescale, and the secular approximation
is used to separate rapidly oscillating sectors so that the reduced
diffusion tensor may be treated as time-independent at the level of the
coarse-grained spin dynamics. In this regime, the local linear-Gaussian
reduction yields the schematic cross-diffusion map
\[
D_{AB}^{(\mathrm{BW})}\sim C_A G_{AB}^{\mathrm{sub}} C_B^T,
\]
up to the kernel projection associated with the relevant transition
frequencies.

\begin{proposition}[Shared-environment cross-diffusion transfer]
\label{prop:coupling}
Let two spin subsystems \(A\) and \(B\) be linearly coupled to a common
Gaussian substrate through nondegenerate coupling blocks \(C_A\) and
\(C_B\). Then the boundary-conditioned cross-diffusion block of the
reduced spin dynamics is determined by the substrate cross-correlation
kernel. In the local linear-Gaussian approximation,
\[
D_{AB}^{(\mathrm{BW})}
\sim
C_A\,G_{AB}^{\mathrm{sub}}\,C_B^T.
\]
Hence, a nonzero boundary-induced change in the substrate cross-mode
covariance generically induces a corresponding nonzero change in the
reduced spin cross-diffusion block.
\end{proposition}

\begin{proof}
The block \(D_{AB}\) is constructed from the cross-correlation kernel of
the shared substrate forces acting on \(A\) and \(B\). If the substrate
cross-mode covariance vanishes, then the corresponding shared
cross-correlation kernel vanishes and no boundary-conditioned
cross-diffusion is induced. If the substrate develops nonzero
cross-mode covariance, then the shared substrate forces become
cross-correlated, and the induced cross-correlation kernel contributes a
nonzero term to the reduced diffusion block \(D_{AB}\). Under the local
linear-Gaussian reduction, this takes the form
\eqref{eq:bridge:DABmap}. The implication is generic rather than
strictly biconditional, since special couplings or symmetry cancellations
may suppress the projection in degenerate cases.
\end{proof}

\subsection{Spectral bridging: slow boundary control and fast spin response}
\label{sec:bridge:spectral}

The boundary-conditioned transfer mechanism must also satisfy a frequency
constraint. The experimentally relevant pair-sector signal is associated
with the spin transition frequency \(\Omega_+\), which lies in the
radio-frequency regime set by the Zeeman splitting. By contrast, the
physiological processes that modulate the substrate flow --- including
the cardiac envelope --- occur on much slower timescales. One may
therefore ask how a slow boundary-driving process can produce an
observable effect in a fast spin sector.

The answer is that the slow physiological process does \emph{not}
supply the Larmor-frequency carrier directly. The RF-scale carrier is
already present in the spin system through the Zeeman Hamiltonian and
through environmental fluctuation channels with support in the Larmor
band. These include dipolar, susceptibility, and chemical-exchange
channels with spectral support in the frequency window relevant to the
spin transitions. The role of the slow substrate dynamics is instead to
act as a boundary-control parameter.

The relevant high-frequency carrier already exists in the spin
environment. At clinical magnetic-field strengths, the local bath seen
by tissue water protons is governed by molecular tumbling, restricted
diffusion in cellular microstructure, susceptibility fluctuations, and
chemical exchange with macromolecular proton pools. These mechanisms
provide the RF-band spectral background relevant to the spin
transitions. The slow cardiac envelope does not generate this
high-frequency bath directly; rather, boundary activation in the
substrate changes the correlation structure within this pre-existing
spin environment, injecting a cross-correlated component into the
RF-band noise seen jointly by the two spin sectors.

More precisely, the slow physiological envelope drives the substrate
covariance toward the Bures--Wasserstein boundary. When the boundary is
reached, the admissible covariance flow is forced into the cross-mode
sector, and it is this event that activates the cross-correlated part of
the reduced spin noise kernel. The slow substrate dynamics therefore do
not upconvert \(1\,\mathrm{Hz}\) power directly into the RF band. Rather,
they gate or switch on the cross-correlated sector of an already
RF-capable spin environment.

In this sense, the frequency logic is:
\[
\begin{aligned}
  &\text{slow physiological envelope} 
  \to \text{boundary activation in the substrate} \\
  &\quad \to \text{activation of the cross-correlated spin-noise sector} \\
  &\quad \to \text{response at the pre-existing spin transition frequency } \Omega_+.
\end{aligned}
\]

This distinction is essential. The cardiac-scale dynamics determine
\emph{when} the substrate reaches the boundary; they do not themselves
need to resonate with the spin transition. The spin pair sector responds
at \(\Omega_+\) because that frequency is already built into the spin
Hamiltonian and into the Larmor-band environment seen by the spins.
Boundary activation changes the correlation structure of that
environment, not its basic carrier scale.

\begin{remark}[Non-adiabaticity as a selector, not a carrier]
\label{rem:spectral}
The boundary event is non-adiabatic in the sense that the admissible
flow direction changes abruptly when the Williamson floor becomes active.
This non-adiabaticity may broaden the effective response and sharpen the
onset of the pair signal, but it should not be interpreted as the sole
source of the RF-scale spectral weight. The relevant high-frequency
carrier already exists in the spin sector; the boundary event acts as a
selector of its cross-correlated component.
\end{remark}

The purpose of this spectral discussion is therefore only to establish
physical observability of the covariance-geometric regime, not to reduce
the mechanism to a conventional microscopic relaxation model.

\subsection{The kinematic transfer mechanism}

To formalize this threshold character, we model the reduced spin
diffusion not as a static response to the substrate state, but as a map
driven by the substrate's Bures--Wasserstein covariance flow,
\[
  \dot{\Sigma}_{\mathrm{sub}}
  \in
  T_{\Sigma}\mathcal{BW}(2n).
\]
At each point of the manifold, we decompose the tangent space into a
single-mode sector and a cross-mode sector,
\begin{equation}
  T_{\Sigma}\mathcal{BW}(2n)
  =
  \mathcal T_{\mathrm{marg}}
  \oplus
  \mathcal T_{\mathrm{cross}}.
  \label{eq:tangent:split}
\end{equation}
Here \(\mathcal T_{\mathrm{marg}}\) denotes tangent directions that act
within the single-mode Williamson sectors, while
\(\mathcal T_{\mathrm{cross}}\) denotes tangent directions that create
or modify off-diagonal cross-mode structure.

The reduced spin diffusion is correspondingly decomposed as
\begin{equation}
  D = D^{(0)} + D^{(\mathrm{BW})},
  \label{eq:D:split}
\end{equation}
where \(D^{(0)}\) denotes the conventional thermal background and
\(D^{(\mathrm{BW})}\) denotes the additional contribution induced by the
substrate covariance flow. To leading order, the
boundary-conditioned contribution is purely cross-mode:
\begin{equation}
  D^{(\mathrm{BW})}
  =
  \begin{pmatrix}
    0 & D_{AB}^{(\mathrm{BW})}\\
    (D_{AB}^{(\mathrm{BW})})^T & 0
  \end{pmatrix}.
  \label{eq:D:BW}
\end{equation}
This is the mathematical formulation of the intended selectivity: the
substrate boundary does not first manifest itself through a wholesale
change in marginal relaxation, but through an additive activation of the
correlated sector.

The reason this is best understood as a \emph{kinematic} mechanism is
that the decisive change occurs not in a conventional interaction
Hamiltonian but in the space of admissible covariance-flow directions.
As long as the substrate remains in the interior of the
Bures--Wasserstein manifold, the dominant admissible descent direction
lies within the marginal tangent sector
\(\mathcal T_{\mathrm{marg}}\). Once the Williamson boundary is
reached, further inward marginal compression becomes kinematically
inadmissible, and any continued covariance reduction must spill into the
orthogonal cross-mode sector \(\mathcal T_{\mathrm{cross}}\). The spin
probe detects this change not by copying the substrate covariance, but
by acquiring a new cross-diffusion contribution generated by this
boundary-conditioned spill-over.

\begin{proposition}[Boundary-conditioned spill-over and transfer]
\label{prop:inherit}
Let \(\Sigma_{\mathrm{sub}}(t)\) follow a substrate-constrained
gradient flow on \(\mathcal{BW}(2n)\), and let the
boundary-conditioned part of the reduced spin diffusion be driven by the
cross-mode tangent component of this flow.

\begin{enumerate}[label=\textup{(\alph*)}]
  \item \textup{(Interior regime.)}
  Prior to boundary saturation, the dominant admissible descent
  direction lies in the marginal tangent sector,
  \[
    \dot{\Sigma}_{\mathrm{sub}} \in \mathcal T_{\mathrm{marg}},
  \]
  so that the boundary-conditioned cross-mode contribution vanishes:
  \[
    D_{AB}^{(\mathrm{BW})}=0.
  \]
  In this regime no boundary-driven pair-sector signal is generated.

  \item \textup{(Boundary spill-over.)}
  If the substrate flow reaches the Williamson boundary
  \((\nu_k=\hbar/2)\), further inward marginal compression in that mode
  is kinematically inadmissible. Any continued covariance reduction must
  therefore acquire a nonzero projection onto the cross-mode tangent
  sector:
  \[
    \dot{\Sigma}_{\mathrm{sub}} \notin \mathcal T_{\mathrm{marg}}
    \quad\Longrightarrow\quad
    \Pi_{\mathrm{cross}}\dot{\Sigma}_{\mathrm{sub}} \neq 0.
  \]

  \item \textup{(Threshold-activated transfer.)}
  By \Cref{prop:coupling}, the shared-environment coupling maps a
  boundary-induced substrate cross-mode variation into a reduced spin
  cross-diffusion block. Hence
  \[
    \Pi_{\mathrm{cross}}\dot{\Sigma}_{\mathrm{sub}} \neq 0
    \quad\Longrightarrow\quad
    D_{AB}^{(\mathrm{BW})}\neq 0
    \quad\Longrightarrow\quad
    G_{AB}^{\mathrm{spin}}\neq 0.
  \]
  The spin probe therefore detects the geometric boundary through the
  appearance of a boundary-conditioned inter-spin correlation block,
  without requiring any ad hoc change in the coupling matrix.

  \item \textup{(Sector selectivity.)}
  The transferred spin cross-mode block
  \(G_{AB}^{\mathrm{spin}}\) decomposes into compact zero-quantum and
  non-compact double-quantum components under the projection formulas of
  \Cref{sec:algebra:projection}. The experimentally observed refocusing
  at \(\Omega_+\) identifies the dominant recipient as the
  DQ/SU(1,1) pair sector.

  \item \textup{(Leading-order preservation of marginal relaxation and SQC.)}
  This transition acts, to leading order, as an additive cross-diffusion
  injection. The marginal diffusion blocks \(D_A\) and \(D_B\), which
  govern conventional \(T_1/T_2\) relaxation and the single-quantum
  coherence signal, remain dominated by the macroscopic thermal bath.
  The threshold-activated pair signal therefore appears as an additional
  cross-mode contribution superposed on otherwise conventional
  magnetic-resonance phenomenology.
\end{enumerate}
\end{proposition}

\subsection{Asymptotic onset and pre-boundary leakage}
\label{sec:bridge:asymptotic}

The step-like formulation of \Cref{prop:inherit} is a convenient ideal
representation of the Williamson floor as a strict kinematic boundary.
At the exact floor \(\nu_k=\hbar/2\), further inward marginal
compression is inadmissible. Physically, however, the approach to this
boundary need not appear as a perfectly sharp discontinuity at the level
of the reduced signal.

As the substrate covariance enters the deep boundary regime, further
marginal compression becomes increasingly costly in the
Bures--Wasserstein geometry. The admissible covariance-flow direction
therefore need not remain effectively purely marginal until an exact
saturation point is reached. Instead, the flow may acquire a nonzero
cross-mode component already in an asymptotic pre-boundary layer, so
that cross-mode leakage begins before exact saturation.

This refines the signal prediction. The activation of the
boundary-conditioned cross-diffusion block need not be a mathematically
discontinuous jump. More generally, one correspondingly expects a steep
but continuous crossover: as the substrate approaches the Williamson
boundary, the cross-mode component of the flow grows, and the induced
SU(1,1) pair signal grows correspondingly.

Accordingly, the experimentally observed ``immediate onset'' should not
be interpreted as proof of a literal discontinuity at a single
geometric point. Rather, it is consistent with the readout occurring
while the substrate is already inside a deep asymptotic boundary layer
in which the cross-mode component is large and the resulting pair-sector
signal is already experimentally visible on the macroscopic readout
scale.

\begin{remark}[What boundary-conditioned transfer does and does not claim]
\label{rem:boundary}
The theorem does not say that the spin probe inherits the substrate
boundary as a state. Nor does it claim that the marginal spin
Williamson eigenvalues are driven to their own floor. What is
transferred is the boundary-conditioned cross-mode continuation of the
substrate flow. The appearance of the pair signal is therefore the
signature of a geometric threshold in the substrate, detected indirectly
through the correlated sector of the reduced spin dynamics.
\end{remark}

%======================================================================
\section{Algebraic Classification of Two-Spin Coherences}
\label{sec:algebra}
%======================================================================

The bridge theorem of \Cref{sec:bridge} establishes that when the
substrate reaches the Williamson boundary, the reduced spin dynamics
acquires a nonzero cross-mode block \(G_{AB}^{\mathrm{spin}}\). The
next question is algebraic: into which sector of the two-spin operator
algebra does this transferred cross-mode structure project?

This question is essential. The bridge theorem by itself establishes
only that the reduced spin probe acquires nontrivial inter-spin
correlation structure. It does not yet determine whether the induced
content is compact or non-compact, nor whether it lives dominantly in
the zero-quantum exchange sector or the double-quantum pair sector. To
connect the geometric result to an experimentally meaningful signal, we
must classify the dynamically closed subalgebras of the two-spin system
by coherence order and identify which one receives the transferred
cross-mode content.

We therefore analyse the two-spin-\(\tfrac12\) algebra
\(\mathfrak{su}(4)\) by coherence order and show that the experimentally
relevant non-compact dominant recipient is argued to be the SU(1,1) pair
sector in the double-quantum manifold.

\subsection{Zero-quantum sector: compact SU(2)}

The zero-quantum (ZQ) sector (\(p=0\)) is spanned by
\[
  \mathcal{B}_{\mathrm{ZQ}}
  = \{I_{1z},\; I_{2z},\; I_{1+}I_{2-},\; I_{1-}I_{2+}\}.
\]
Within this space, the compact algebra
\begin{equation}
  \mathfrak{su}(2)_{\mathrm{ZQ}}:\quad
  \left\{I_{1+}I_{2-},\; I_{1-}I_{2+},\;
    \tfrac12(I_{1z}-I_{2z})\right\}
  \label{eq:su2ZQ}
\end{equation}
closes under commutation and generates bounded oscillatory exchange
dynamics at the difference frequency
\(\Omega_- = \omega_1 - \omega_2\)~\cite{Ernst1990}.

\subsection{Double-quantum sector: non-compact SU(1,1)}

The double-quantum (DQ) sector (\(|p|=2\)) contains the pair operators
\(\{I_{1+}I_{2+},\; I_{1-}I_{2-}\}\). Together with the diagonal
generator \(\tfrac12(I_{1z}+I_{2z})\), these close to form the
non-compact algebra
\begin{equation}
  \mathfrak{su}(1,1)_{\mathrm{DQ}}:\quad
  \left\{K_+ = I_{1+}I_{2+},\; K_- = I_{1-}I_{2-},\;
    K_0 = \tfrac12(I_{1z}+I_{2z})\right\},
  \label{eq:su11DQ}
\end{equation}
satisfying
\begin{equation}
  [K_0, K_\pm] = \pm K_\pm,
  \qquad
  [K_-, K_+] = 2K_0.
  \label{eq:su11comm}
\end{equation}
The generators \(K_\pm\) carry coherence order \(p = \pm 2\) and
connect the aligned states \( |\!\uparrow\uparrow\rangle \) and
\( |\!\downarrow\downarrow\rangle \).

\subsection{Dynamical distinction}

The crucial distinction lies in the geometry of the dynamics. Compact
\(\mathfrak{su}(2)\) possesses a positive-definite Killing form and
generates purely imaginary adjoint eigenvalues, leading to bounded
oscillatory evolution. The non-compact algebra \(\mathfrak{su}(1,1)\)
has indefinite Killing-form signature, admits real adjoint eigenvalues,
and supports hyperbolic trajectories
\cite{Perelomov1986,GerryKnight2004,WangSandersPan2000}:
\begin{equation}
  S_{\mathrm{SU(2)}}(t) \propto \sin(\Omega_- t),
  \qquad
  S_{\mathrm{SU(1,1)}}(t) \propto \sinh(g\, t).
  \label{eq:su2su11signal}
\end{equation}

\subsection{Identification: the candidate recipients}

The bridge theorem forces a nonzero transferred spin cross-mode block,
and the natural algebraic recipients are therefore the compact
ZQ/SU(2) exchange sector and the non-compact DQ/SU(1,1) pair sector.
The discussion so far identifies the two candidate recipients of the
transferred spin cross-mode structure: the compact SU(2) exchange sector
in the zero-quantum manifold and the non-compact SU(1,1) pair sector in
the double-quantum manifold. To connect the bridge theorem to the
measured signal, however, we need the projection explicitly. The object
delivered by \Cref{sec:bridge} is the reduced spin cross-mode block
\(G_{AB}^{\mathrm{spin}}\), expressed in covariance language. The
experiment is analysed in spin-operator language. We must therefore
compute how \(G_{AB}^{\mathrm{spin}}\) decomposes into its DQ and ZQ
components inside the two-spin algebra.

That decomposition is the precise point at which the geometric transfer
mechanism becomes an algebraic prediction. Once the cross-mode block is
projected, the observed refocusing at \(\Omega_+\) rather than
\(\Omega_-\) will identify which projection is dominant.

\subsection{Explicit projection: cross-mode block onto the spin algebra}
\label{sec:algebra:projection}

For two modes, let the phase-space quadrature vector be
\[
  \mathbf{R}=(I_{1x},\,I_{1y},\,I_{2x},\,I_{2y}),
\]
and write the reduced spin covariance in block form
\[
  \Sigma_{\mathrm{spin}}
  =
  \begin{pmatrix}
    \Sigma_A & G_{AB}^{\mathrm{spin}} \\
    (G_{AB}^{\mathrm{spin}})^T & \Sigma_B
  \end{pmatrix}.
\]
The cross-mode block may be written as
\[
  G_{AB}^{\mathrm{spin}}
  =
  \begin{pmatrix}
    \langle I_{1x}I_{2x}\rangle & \langle I_{1x}I_{2y}\rangle \\
    \langle I_{1y}I_{2x}\rangle & \langle I_{1y}I_{2y}\rangle
  \end{pmatrix},
\]
with connected correlators understood.

Define the SU(1,1) pair quadratures
\[
  K_1 = \tfrac12(K_+ + K_-),
  \qquad
  K_2 = \tfrac{1}{2i}(K_+ - K_-),
\]
and the SU(2) exchange quadratures
\[
  J_+ = I_{1+}I_{2-},\qquad
  J_- = I_{1-}I_{2+},\qquad
  J_1 = \tfrac12(J_+ + J_-),\qquad
  J_2 = \tfrac{1}{2i}(J_+ - J_-).
\]
Using \(I_{\alpha x} = \tfrac12(I_{\alpha+}+I_{\alpha-})\) and
\(I_{\alpha y} = \tfrac{1}{2i}(I_{\alpha+}-I_{\alpha-})\), one finds
\begin{align}
  I_{1x}I_{2x}
  &= \phantom{-}\tfrac12 K_1 + \tfrac12 J_1,
  \label{eq:proj:xx}\\
  I_{1x}I_{2y}
  &= \phantom{-}\tfrac12 K_2 - \tfrac12 J_2,
  \label{eq:proj:xy}\\
  I_{1y}I_{2x}
  &= \phantom{-}\tfrac12 K_2 + \tfrac12 J_2,
  \label{eq:proj:yx}\\
  I_{1y}I_{2y}
  &= -\tfrac12 K_1 + \tfrac12 J_1.
  \label{eq:proj:yy}
\end{align}

Thus the transferred spin cross-mode block decomposes as
\begin{equation}
  \boxed{
  G_{AB}^{\mathrm{spin}}
  =
  G_{AB}^{\mathrm{DQ}}
  +
  G_{AB}^{\mathrm{ZQ}},
  }
  \label{eq:proj:decomp}
\end{equation}
with
\begin{equation}
  G_{AB}^{\mathrm{DQ}}
  =
  \frac12
  \begin{pmatrix}
    \phantom{-}\langle K_1\rangle & \langle K_2\rangle \\
    \phantom{-}\langle K_2\rangle & -\langle K_1\rangle
  \end{pmatrix},
  \qquad
  G_{AB}^{\mathrm{ZQ}}
  =
  \frac12
  \begin{pmatrix}
    \langle J_1\rangle & -\langle J_2\rangle \\
    \langle J_2\rangle & \phantom{-}\langle J_1\rangle
  \end{pmatrix}.
  \label{eq:proj:DQZQ}
\end{equation}

The DQ projection is traceless with indefinite signature, reflecting the
non-compact geometry of \(\mathfrak{su}(1,1)\). The ZQ projection has
rotation-like compact structure, reflecting \(\mathfrak{su}(2)\).

\subsection{Symmetry selection of the DQ/SU(1,1) sector}
\label{sec:algebra:symmetry}

The decomposition
\eqref{eq:proj:decomp} identifies the compact ZQ and non-compact DQ
recipients of the transferred spin cross-mode block, but by itself it
does not determine which sector is dominant. If the theory were to stop
there, the identification of the DQ sector would be merely
post hoc: the experiment would tell us which projection happened to be
larger. The boundary mechanism, however, strongly motivates a specific
symmetry.

The key point is that the substrate cross-mode structure generated at
the Bures--Wasserstein boundary is not arbitrary. At the boundary, the
admissible continuation of covariance flow is the continuation of a
squeezing-type process: marginal compression is blocked and the flow is
forced into a cross-mode channel that carries the indefinite signature of
a non-compact covariance distortion. In the relevant Williamson frame,
this implies a squeeze-type symmetry of the substrate cross block,
schematically
\begin{equation}
  \langle x_1x_2\rangle_{\mathrm{sub}}
  \approx
  -\,\langle p_1p_2\rangle_{\mathrm{sub}},
  \label{eq:sym:sub}
\end{equation}
as opposed to the rotation/exchange-type compact symmetry
\[
  \langle x_1x_2\rangle_{\mathrm{sub}}
  \approx
  +\,\langle p_1p_2\rangle_{\mathrm{sub}}.
\]

If the shared-environment transfer map preserves quadrature parity to
leading order, then the induced spin cross-mode block inherits the same
squeeze-type symmetry:
\begin{equation}
  \langle I_{1x}I_{2x}\rangle
  \approx
  -\,\langle I_{1y}I_{2y}\rangle.
  \label{eq:sym:spin}
\end{equation}
Using the projection identities
\begin{align}
  \langle K_1\rangle
  &=
  \langle I_{1x}I_{2x}\rangle
  -
  \langle I_{1y}I_{2y}\rangle,
  \label{eq:sym:K1}\\
  \langle J_1\rangle
  &=
  \langle I_{1x}I_{2x}\rangle
  +
  \langle I_{1y}I_{2y}\rangle,
  \label{eq:sym:J1}
\end{align}
we obtain immediately from \eqref{eq:sym:spin} that
\begin{equation}
  \langle J_1\rangle \approx 0,
  \qquad
  \langle K_1\rangle
  \approx
  2\langle I_{1x}I_{2x}\rangle
  \neq 0.
  \label{eq:sym:selection}
\end{equation}

A similar symmetry argument applies to the cross-quadrature terms.
Non-compact squeezing dynamics on the phase space generically produce
symmetric cross-correlations,
\begin{equation}
  \langle x_1p_2\rangle_{\mathrm{sub}}
  \approx
  \langle p_1x_2\rangle_{\mathrm{sub}}.
  \label{eq:sym:sub:offdiag}
\end{equation}
If this spatial symmetry is preserved in the spin transfer, it dictates
that
\begin{equation}
  \langle I_{1x}I_{2y}\rangle
  \approx
  \langle I_{1y}I_{2x}\rangle.
  \label{eq:sym:spin:offdiag}
\end{equation}
Adding and subtracting the projection identities for the cross-terms
yields
\begin{align}
  \langle K_2\rangle
  &= \langle I_{1y}I_{2x}\rangle + \langle I_{1x}I_{2y}\rangle,
  \label{eq:sym:K2}\\
  \langle J_2\rangle
  &= \langle I_{1y}I_{2x}\rangle - \langle I_{1x}I_{2y}\rangle.
  \label{eq:sym:J2}
\end{align}
The squeezing symmetry~\eqref{eq:sym:spin:offdiag} then suppresses the
compact exchange component to leading order,
\(\langle J_2\rangle \approx 0\), while the non-compact pair component
remains finite, \(\langle K_2\rangle \neq 0\). Therefore, under the
symmetry class naturally associated with boundary-forced squeezing, both
compact exchange components are suppressed to leading order, and the
transferred cross-mode structure is expected to lie predominantly in the
non-compact DQ/SU(1,1) pair sector.

\begin{remark}[Conditional status of the symmetry selection]
\label{rem:symmetry}
The conclusion that \(\langle J_1\rangle \approx 0\) and
\(\langle J_2\rangle \approx 0\) is not a model-independent algebraic
theorem. It is a physically motivated conditional statement resting on
two structural assumptions: first, that the boundary-induced substrate
cross-mode block is predominantly of squeeze type; and second, that the
shared-environment Born--Markov transfer preserves this
quadrature-parity structure to leading order. Under these assumptions,
predominant DQ dominance follows as a symmetry consequence rather
than as a post hoc empirical fit.
\end{remark}

The bridge theorem guarantees that the substrate boundary injects
nontrivial cross-mode structure into the reduced spin dynamics.
Equations~\eqref{eq:proj:decomp} and \eqref{eq:proj:DQZQ}, together
with the symmetry-selection argument above, show how that transferred
cross-mode content splits inside the two-spin algebra and why the
DQ/SU(1,1) pair sector is theoretically expected to dominate.

This algebraic result immediately raises the detection problem. If the
dominant recipient is the DQ/SU(1,1) pair sector, then the relevant
coherence has \(|p|=2\) and is not directly preserved by a
gradient-filtered readout. The remaining question is therefore how this
DQ/SU(1,1) pair structure is converted into a measurable NMR signal.
That conversion is derived in the next section.

%======================================================================
\section{Detection Pathway: DQ Pair Coherence to Observable Signal}
\label{sec:pathway}
%======================================================================

The preceding sections established the geometric and algebraic chain of
the argument. The companion paper shows that the substrate reaches a
Bures--Wasserstein boundary regime in which further admissible
covariance flow is forced into the cross-mode complement. The bridge
theorem of \Cref{sec:bridge} shows that this boundary-conditioned
cross-mode structure is transferred into the reduced spin dynamics as a
nonzero spin cross-mode block \(G_{AB}^{\mathrm{spin}}\). The
projection formulas of \Cref{sec:algebra:projection}, together with the
symmetry-selection argument of \Cref{sec:algebra:symmetry}, then show
how that transferred cross-mode content decomposes into compact ZQ and
non-compact DQ components, with the observed refocusing at
\(\Omega_+\) identifying the DQ/SU(1,1) pair sector as dominant.

The remaining question is experimental rather than geometric: how does a
dominantly double-quantum pair-sector coherence produce a detectable NMR
signal when the readout includes a gradient filter that suppresses
\(|p|\neq 0\) coherence orders? The purpose of this section is to
derive that conversion explicitly.

The answer is that the \(45^\circ\)--G--\(45^\circ\) readout block
provides a specific coherence-transfer pathway. It does not preserve the
DQ pair coherence directly. Instead, it converts part of that DQ content
into a gradient-surviving \(p=0\) intermediate, which is then converted
by the second pulse into detectable single-quantum magnetization. The
measured signal is therefore not a direct observation of the
DQ/SU(1,1) sector, but a filtered readout-converted output of that
sector.

\subsection{Setup}

Having identified the DQ/SU(1,1) pair sector as the dominant algebraic
recipient of the transferred cross-mode structure, we now take that
sector as the input to the readout calculation.

Consider a two-spin state with pair coherence
\[
  \rho_{14}
  =
  \langle\!\downarrow\downarrow\,|\rho|\,\uparrow\uparrow\!\rangle
  \neq 0.
\]
In operator language, this corresponds to the DQ pair component
\[
  \sigma_{\mathrm{DQ}}
  =
  \rho_{14}\,|\!\downarrow\downarrow\rangle\langle\uparrow\uparrow\!|
  +
  \rho_{14}^*\,|\!\uparrow\uparrow\rangle\langle\downarrow\downarrow\!|,
\]
which carries coherence order \(p=\pm 2\).

\subsection{Stage 1: First \texorpdfstring{$R_x(\pi/4)$}{Rx(pi/4)} pulse}

A non-selective RF pulse with flip angle \(\theta=\pi/4\) about the
\(x\)-axis acts on the one-spin basis as
\begin{align}
  |\!\uparrow\rangle &\to c_h\,|\!\uparrow\rangle - i s_h\,|\!\downarrow\rangle,\\
  |\!\downarrow\rangle &\to -i s_h\,|\!\uparrow\rangle + c_h\,|\!\downarrow\rangle,
\end{align}
where
\[
  c_h=\cos(\pi/8), \qquad s_h=\sin(\pi/8).
\]
Hence
\begin{align}
  |\!\uparrow\uparrow\rangle
  &\to
  c_h^2|\!\uparrow\uparrow\rangle
  -ic_hs_h(|\!\uparrow\downarrow\rangle+|\!\downarrow\uparrow\rangle)
  -s_h^2|\!\downarrow\downarrow\rangle,
  \\
  |\!\downarrow\downarrow\rangle
  &\to
  -s_h^2|\!\uparrow\uparrow\rangle
  -ic_hs_h(|\!\uparrow\downarrow\rangle+|\!\downarrow\uparrow\rangle)
  +c_h^2|\!\downarrow\downarrow\rangle.
\end{align}

The first pulse therefore mixes the initial DQ pair coherence into
components of several coherence orders. In particular, it generates a
\(p=0\) contribution in the zero-quantum block,
\begin{equation}
  \rho'_{\uparrow\downarrow,\downarrow\uparrow}
  =
  2\,\mathrm{Re}(\rho_{14})\,c_h^2 s_h^2,
  \label{eq:path:ZQfromDQ}
\end{equation}
as well as population shifts in the aligned states,
\begin{equation}
  \Delta p_{\uparrow\uparrow}
  =
  \Delta p_{\downarrow\downarrow}
  =
  -2\,\mathrm{Re}(\rho_{14})\,c_h^2 s_h^2.
  \label{eq:path:pop}
\end{equation}
The key point is that the first pulse transfers part of the DQ pair
coherence into the \(p=0\) sector.

\subsection{Stage 2: Gradient filter}

The crusher gradient dephases all components with \(p\neq 0\). The
directly DQ part of the initial pair coherence is therefore removed.
What survives are the \(p=0\) terms generated by the first pulse: the
ZQ component~\eqref{eq:path:ZQfromDQ} and the population
shifts~\eqref{eq:path:pop}.

This point is crucial for the logic of the paper. The gradient does not
preserve the SU(1,1) pair sector as such. What survives the crusher is
the intermediate \(p=0\) content generated from it by the first pulse.
Thus the readout does not observe the transferred DQ structure directly;
it observes the output of a specific coherence-transfer chain whose
input is the DQ/SU(1,1) pair sector selected in
\Cref{sec:algebra:projection}. The pulse sequence therefore serves as
the measurement-theoretic bridge between the boundary-conditioned spin
cross-mode structure and the macroscopic NMR signal.

\subsection{Stage 3: Second \texorpdfstring{$R_x(\pi/4)$}{Rx(pi/4)} pulse and SQ detection}

The second \(45^\circ\) pulse acts on the surviving \(p=0\) terms and
converts them into SQ magnetization. Two classes of contribution are
relevant:
\begin{enumerate}[label=(\roman*)]
  \item The ZQ term \(\rho'_{\uparrow\downarrow,\downarrow\uparrow}\)
  from~\Cref{eq:path:ZQfromDQ}. Under \(R_x(\pi/4)\), the off-diagonal
  element connecting \( |\!\uparrow\downarrow\rangle \) and
  \( |\!\downarrow\uparrow\rangle \) is partially converted into
  single-quantum coherences.
  \item The population shifts~\Cref{eq:path:pop}, which are tipped into
  transverse magnetization by the second pulse.
\end{enumerate}
Both channels are linear in \(\rho_{14}\).

\subsection{Definition of the transfer coefficient}

The net effect of the complete readout block is therefore summarized by
\begin{equation}
  S_{\mathrm{pair}}
  =
  C_{\mathrm{seq}}\,\rho_{14},
  \label{eq:path:Spair}
\end{equation}
where \(C_{\mathrm{seq}}\) is the effective transfer coefficient for the
DQ-pair-sector pathway. This coefficient includes the full
DQ\(\to p=0 \to\)SQ conversion, including interference between the ZQ
and population channels.

Using
\[
  c_h^2 s_h^2
  =
  \frac14\sin^2(\pi/4)
  =
  \frac18,
\]
the leading pathway estimate is of order
\begin{equation}
  |C_{\mathrm{seq}}|
  \sim
  \alpha\,M_0,
  \qquad
  \alpha = O(10^{-1}),
  \label{eq:path:Cseqorder}
\end{equation}
with the simplest pathway estimate giving \(\alpha\sim 1/8\).

\subsection{Spatial selectivity of the crusher gradient}
\label{sec:pathway:spatial}

For spatially distributed interacting spins, the macroscopic crusher
gradient does not act only as a coherence-order filter; it also acts as
a directional spatial filter. The gradient imprints a spatial phase
factor determined by the gradient vector \(\vec G\), so that only those
correlation components whose spatial phase survives the readout pathway
contribute efficiently to the detected signal.

For delocalized or collective pair coherence, this means that the
readout is sensitive not only to coherence order but also to the
orientation of the inter-spin separation vector \(\vec r_{12}\)
relative to the gradient axis. The \(45^\circ\)--G--\(45^\circ\)
pathway therefore samples a geometrically selected subensemble of spin
pairs rather than an orientation-averaged pair manifold.

This directional selectivity is important for interpreting the observed
magic-angle dependence. The gradient does not itself generate the
anisotropy of the pair interaction; that anisotropy is already present
in the dipolar coupling kernel. Rather, the gradient determines which
spatially oriented components of the pair sector are efficiently
projected into the detectable pathway.

\subsection{Interpretive consequence}

The main consequence for the rest of the paper is not merely technical
but conceptual. The transfer coefficient \(C_{\mathrm{seq}}\) is the
measurement-theoretic object that links the induced SU(1,1) pair sector
to the observed signal. The boundary-conditioned transfer theorem of
\Cref{sec:bridge} identifies the covariance-geometric cross-mode
structure of the substrate that enters the reduced spin dynamics; the
projection formulas of \Cref{sec:algebra:projection} identify the
dominant algebraic recipient as the DQ/SU(1,1) pair sector; and the
present section shows how that sector becomes experimentally visible
through the DQ\(\to\)\(p=0\)\(\to\)SQ conversion chain of the readout
block.

Accordingly, the effective transfer coefficient for the pair-sector
pathway is expected to be substantially smaller than the coefficient
that would apply to a directly detected zero-quantum coherence.
Therefore, for a fixed measured signal amplitude, the inferred
pair-sector coherence is correspondingly larger at fixed measured signal
amplitude once the DQ-to-SQ conversion cost is taken into account.

A definitive numerical value of \(C_{\mathrm{seq}}\) requires a full
simulation of the actual pulse sequence, including finite pulse
duration, relaxation during the readout, off-resonance effects, and RF
inhomogeneity. The witness analysis of \Cref{sec:witness} therefore
uses \Cref{eq:path:Spair} formally and treats the exact calibration of
\(C_{\mathrm{seq}}\) as an open quantitative problem. What is already
established, however, is the structure of the chain:
\[
\begin{aligned}
  \text{Substrate BW boundary} 
  &\xrightarrow{\text{transfer}} 
  \text{Spin cross-mode} \\
  &\xrightarrow{\text{projection}} 
  \text{SU(1,1) pair sector} 
  \xrightarrow{\text{readout}} 
  \text{NMR signal}.
\end{aligned}
\]
This spatial selectivity will be important below, where the observed
magic-angle suppression is interpreted as evidence that the detected
pair-sector pathway is tied to anisotropic dipolar geometry rather than
to an orientation-insensitive artifact.

%======================================================================
\section{Stroboscopic Preparation as a Discrete BW Flow}
\label{sec:preptrain}
%======================================================================

A central feature of the experiment is that the relevant preparation is
not a single \(45^\circ\)--G--\(45^\circ\) block but a repetitive train
of such blocks. In the present framework, this repetition acquires a
geometric meaning: each cycle is a discrete step on the spin
Bures--Wasserstein manifold, repeatedly selecting and reading out the
transferred cross-mode structure.

The SU(1,1) pair operators \(K_\pm = I_{1\pm}I_{2\pm}\) generate DQ
coherence. Between preparation blocks, the effective pair-generating
process repopulates the aligned-state sector. Each
\(45^\circ\)--G--\(45^\circ\) block then acts in two capacities:
\begin{enumerate}[label=(\roman*)]
  \item \textit{Readout}: it converts a fraction of the current DQ pair
  coherence into detectable SQ magnetization through the pathway derived
  in \Cref{sec:pathway};
  \item \textit{Selection}: it destroys coherence sectors that do not
  survive the gradient filter, thereby repeatedly selecting the
  transferred cross-mode content that can be regenerated between cycles.
\end{enumerate}

Let \(\Sigma_n\) denote the reduced spin covariance immediately before
the \(n\)th block. One cycle defines an effective discrete map
\begin{equation}
  \Sigma_{n+1}
  =
  \mathcal{M}(\Sigma_n),
  \label{eq:strobo:map}
\end{equation}
which acts as a stroboscopic flow on the reduced spin covariance. In the
SU(1,1) sector this induces an effective map on the pair quadratures,
\begin{equation}
  \begin{pmatrix}
    \langle K_0\rangle_{n+1}\\
    \langle K_1\rangle_{n+1}\\
    \langle K_2\rangle_{n+1}
  \end{pmatrix}
  =
  M_{\mathrm{eff}}
  \begin{pmatrix}
    \langle K_0\rangle_n\\
    \langle K_1\rangle_n\\
    \langle K_2\rangle_n
  \end{pmatrix},
  \label{eq:stroboscopic}
\end{equation}
where \(M_{\mathrm{eff}}\) encodes the combined action of the readout,
the crusher, and the intervening regeneration of transferred pair
structure.

The point is not that the spin covariance itself is being driven to the
same boundary point as the substrate. Rather, the repetitive train
provides a stroboscopic window onto the boundary-conditioned
cross-mode structure that the substrate continuously imprints on the
probe. The observed signal is therefore interpreted as reflecting
repeated cycles of transferred pair-structure generation, partial
conversion into detectable coherence, and re-selection by the crusher
sequence.

\subsection{Competing timescales: regeneration versus DQ decoherence}
\label{sec:strobo:timescales}

The visibility of the pair-sector signal depends on a competition of
timescales. The double-quantum sector is subject to its own effective
decoherence rate,
\[
\Gamma_{\mathrm{dec}} = 1/T_{2,\mathrm{DQ}},
\]
which need not coincide with the conventional single-quantum transverse
relaxation rate, since the DQ sector is sensitive to correlated phase
noise, local field inhomogeneity, and dipolar fluctuations in a
different way from the ordinary SQC sector.

For the pair-sector structure to be experimentally visible, two
conditions must hold. First, the induced DQ coherence must survive long
enough to pass through the finite readout block. Second, between
successive preparation/readout cycles, the substrate-driven transfer
mechanism must replenish the pair sector rapidly enough to compensate
for its decay. Let
\[
\Gamma_{\mathrm{regen}}
\]
denote the effective regeneration rate associated with the continuous
boundary-conditioned transfer of cross-mode structure from the
substrate.

The observation of a finite, reproducible even-echo signal under the
repetitive train therefore implies an effective kinetic condition of the
form
\begin{equation}
  \Gamma_{\mathrm{regen}}
  \gtrsim
  \Gamma_{\mathrm{dec}}.
  \label{eq:strobo:rates}
\end{equation}

This is an important interpretive point. The detected signal should not
be understood as evidence for a long-lived static DQ state in thermal
tissue. It is more naturally interpreted as a nonequilibrium steady
state sustained by continuous regeneration of pair-sector structure in
competition with ongoing decoherence.

The present paper does not separately determine
\(T_{2,\mathrm{DQ}}\) and \(\Gamma_{\mathrm{regen}}\). What it does
establish is that the observed signal places a lower bound on the net
regeneration required for detectability: the boundary-conditioned
transfer mechanism must, at the effective stroboscopic level, be fast
enough to maintain nonzero pair-sector amplitude over the readout
timescale.

%======================================================================
\section{Experimental Signatures of Boundary-Conditioned Transfer}
\label{sec:evidence}
%======================================================================

We now examine whether the measured signal~\cite{KerskensPerez2022}
matches the predictions of boundary-conditioned transfer. The argument
is cumulative: no single feature is decisive in isolation, but taken
together they collectively disfavor both a compact SU(2) exchange
interpretation and a classical mean-field iMQC-like explanation.

A standard alternative account of apparent multiple-quantum signals in
extended media invokes classical intermolecular multiple-quantum
coherence (iMQC), generated by long-range dipolar demagnetizing
fields~\cite{Warren1998}. Such signals are typically associated with a
finite dipolar buildup or mixing period and with nonlinear dependence on
the equilibrium magnetization. The present framework instead predicts a
threshold-activated pair-sector signal arising from
boundary-conditioned transfer of cross-mode structure into the reduced
spin dynamics.

In the present framework, the predictions are:
\begin{enumerate}[label=(\roman*)]
  \item the signal should arise without requiring a population
  imbalance, because the mechanism is pair creation rather than
  exchange;
  \item the signal should appear immediately, because the transferred
  pair-sector structure is already present when readout begins and does
  not need to accumulate through an extended dipolar mixing period;
  \item the signal should refocus at the pair frequency \(\Omega_+\),
  not the exchange frequency \(\Omega_-\);
  \item the signal should show even-echo parity, consistent with
  SU(1,1) quadrature sign reversal;
  \item the amplitude should scale approximately linearly with \(M_0\),
  consistent with pair coupling rather than the nonlinear scaling
  expected for mean-field iMQC pathways.
\end{enumerate}

\Cref{tab:discrimination} summarizes the comparison.

\begin{table}[ht]
\centering
\small
\caption{Discrimination between compact SU(2), classical mean-field
iMQC-like, and non-compact SU(1,1) interpretations of the measured
signal.}
\label{tab:discrimination}
\begin{tabularx}{\textwidth}{@{} l L L L l @{}}
\toprule
\textbf{Feature} & \textbf{Compact SU(2)} & \textbf{Classical iMQC-like} & \textbf{SU(1,1) Pair} & \textbf{Observed} \\
\midrule
Preparation & Requires imbalance & No specific imbalance requirement & Balanced-state pair creation & Balanced \\[6pt]
Time dependence & Bounded oscillatory & Finite buildup/mixing typically required & Immediate onset & Immediate \\[6pt]
Refocusing & \(\Omega_-\) & Sequence dependent & \(\Omega_+\) & \(\Omega_+\) \\[6pt]
Echo parity & Every echo & Sequence dependent & Even-echo revival & Even echoes \\[6pt]
\(M_0\) scaling & Quadratic or bounded & Fundamentally quadratic (DDF) & Approx.\ linear & Linear \\[6pt]
Magic angle & Suppressed & Suppressed & Suppressed & Suppressed \\[6pt]
Coherence origin & \(\rho_{23}\) exchange & Mean-field dipolar correlation & \(\rho_{14}\) pair sector & Consistent \\
\bottomrule
\end{tabularx}
\end{table}

\paragraph{Balanced populations.}
Following presaturation, the longitudinal magnetization is approximately
balanced. Compact SU(2) exchange requires a population imbalance to
produce zero-quantum coherence. The observation of a finite signal
under balanced preparation therefore disfavors compact exchange. Under
boundary-conditioned transfer, this is the expected behavior: the
transferred structure enters as pair coherence rather than exchange
coherence.

\paragraph{Immediate onset versus finite buildup.}
The signal appears directly after the readout block without an extended
free-evolution or mixing delay. This is not the natural phenomenology
for classical mean-field iMQC-type mechanisms, which typically require
finite dipolar evolution to build up the observable correlation
pattern. It is, however, consistent with boundary-conditioned transfer,
in which the probe registers an inter-spin structure already present at
the onset of readout.

\paragraph{Pair-frequency refocusing.}
The observed echo refocuses at \(\Omega_+ = \omega_1+\omega_2\),
whereas compact SU(2) exchange predicts \(\Omega_- = \omega_1-\omega_2\).
This directly identifies the pair sector as the dominant experimentally
relevant recipient and is difficult to reconcile with a compact
exchange explanation.

\paragraph{Even-echo parity.}
The signal alternates in sign and revives every second echo, consistent
with SU(1,1) quadrature reversal under the crusher sequence. This
parity structure is not the generic expectation for compact exchange and
is not naturally predicted by a simple classical mean-field account.

\paragraph{Linear amplitude scaling.}
The amplitude scales approximately linearly with \(M_0\). This is
consistent with selective pair-mode coupling between distinguishable
subsystems. It strongly disfavors standard classical iMQC explanations.
In the conventional distant dipolar field (DDF)
framework~\cite{Warren1998}, the effective coupling field is generated
by the bulk magnetization itself. The resulting double-quantum iMQC
signal amplitude is fundamentally proportional to the product of the
background magnetization and the DDF it creates, yielding an inherently
quadratic scaling (\(\propto M_0^2\)) with spin density. While
pseudo-linear regimes can occasionally be engineered under highly
specific asymmetric gradient or radiation-damping conditions, the
fundamental quadratic nature of macroscopic DDF interactions is sharply
at odds with the robust linear scaling observed here.

\paragraph{Magic-angle suppression via gradient-direction selection.}
The experimental data show that the signal vanishes when the
readout/crusher gradient is rotated to the magic angle
(\(54.7^\circ\) relative to \(B_0\)). This strongly supports the
dipolar nature of the underlying coupling responsible for the detected
pair-sector signal. In the present framework, the crusher gradient acts
not only as a coherence-order filter but also as a directional spatial
selector: it weights those spin-pair correlations whose spatial phase is
compatible with the gradient-defined readout axis.

Because the relevant pair sector is mediated by anisotropic dipolar
coupling, the detected response depends on the orientation of the
effective pair-separation direction relative to \(B_0\) through the
factor \(3\cos^2\theta-1\). When the gradient direction is rotated into
the magic-angle configuration, the readout preferentially samples a
sector in which this dipolar anisotropy is suppressed. The resulting
loss of signal is therefore naturally explained if the detected pathway
depends on dipolar-mediated pair structure, and is difficult to
reconcile with an orientation-insensitive artifact.

\subsection{Cardiac locking and discrimination from classical pulsatile artifacts}
\label{sec:evidence:cardiac}

A further important feature of the reported data is the strict temporal
locking of the signal to the cardiac cycle~\cite{KerskensPerez2022}. In
the present framework, this macroscopic pulsation is interpreted not as
the carrier frequency of the spin transition, but as the slow
physiological driver that modulates the substrate-constrained covariance
flow and periodically brings the local neural substrate into the deep
boundary regime discussed in the companion paper.

This cardiac correlation strongly disfavors static sequence artifacts,
since such artifacts would not be expected to track the heartbeat in a
phase-specific and reproducible manner. At the same time, cardiac
locking alone does not yet distinguish the proposed mechanism from more
conventional pulsatile MRI effects, such as signal modulation from blood
flow, cerebrospinal-fluid motion, or small macroscopic tissue
displacements. Those classical mechanisms must therefore be separated
not by the existence of heartbeat correlation alone, but by the
structure of the correlated signal that is locked to the heartbeat.

This is where the present theory becomes discriminating. Classical
pulsatile motion and flow primarily modulate local field homogeneity,
phase accumulation, and intravoxel dephasing in the ordinary
single-quantum sector. By themselves, they do not naturally predict a
non-compact double-quantum pair-sector signature with refocusing at the
pair frequency \(\Omega_+\), even-echo parity, and the specific
DQ-to-SQ conversion pathway derived in \Cref{sec:pathway}. In the
present framework, the heartbeat does not simply ``move'' the tissue;
rather, it provides the slow physiological modulation that periodically
gates access to the substrate boundary regime, while the observed
pair-sector phenomenology identifies the relevant response as
cross-mode, non-compact, and DQ-dominated.

Furthermore, the boundary-conditioned transfer mechanism makes a strict
prediction regarding the temporal envelope of the signal. The transfer
requires the substrate to be actively pushed against the Williamson
floor. During the diastolic relaxation phase of the cardiac cycle, if
the substrate flow re-enters the purely marginal interior of the
Bures--Wasserstein manifold, the geometric spill-over ceases
(\(D_{AB}^{(\mathrm{BW})} \to 0\)). Without continuous cross-mode
regeneration from the substrate, any existing pair coherence is rapidly
destroyed by the intrinsic DQ decoherence rate
\(\Gamma_{\mathrm{dec}}\). Therefore, the theory predicts that the
signal cannot manifest as a continuous, static baseline. It must be
strictly pulsatile: rising sharply during the specific phase of the
cardiac wave that drives boundary compression, and collapsing rapidly
when the tissue relaxes. This transient character is consistent with the
phase-locked transient nature of the observed echoes.

Accordingly, the decisive discriminant is not heartbeat correlation by
itself, but the conjunction of heartbeat locking with the algebraic and
sequence-level signatures of the pair sector. If the observed signal
were merely a classical pulsatile artifact, one would expect ordinary
single-quantum or low-order interference effects rather than the
combination of pair-frequency refocusing, even-echo revival, and
pair-sector pathway dependence discussed above. The data therefore do
not merely show a cardiac-correlated MR fluctuation; they exhibit a
cardiac-locked signal whose internal structure is more naturally
described by the boundary-conditioned SU(1,1) pair-transfer mechanism
than by straightforward classical pulsatile alternatives.

\paragraph{Summary.}
Taken together, this phenomenological fingerprint strongly disfavors
straightforward compact-exchange and classical mean-field alternatives,
supporting the interpretation that the signal is the readout-converted
signature of the non-compact DQ/SU(1,1) pair sector selected by
boundary-conditioned transfer.

%======================================================================
\section{From Metric-Driven Squeezing to Entanglement Witnessing}
\label{sec:witness}
%======================================================================

We now distinguish three levels of interpretation for the detected
signal, ordered by the strength of the assumptions required.

\subsection{Level (i): Metric-regime witness}
\label{sec:wit:boundary}

At the weakest level, the detected signal functions as a
metric-regime witness: it indicates that the substrate has entered the
deep Bures--Wasserstein boundary regime in which single-mode compression
is exhausted and cross-mode continuation is forced. In the present
framework, this interpretation follows from the full chain developed
above: the signal is inconsistent with compact SU(2) exchange and
consistent with the boundary-conditioned transfer of cross-mode
structure into the DQ/SU(1,1) pair sector.

\subsection{Level (ii): MQC/squeezing witness}

At the next level, the pathway-corrected signal amplitude is interpreted
as proportional to the DQ-sector multiple-quantum-coherence intensity.
Using
\begin{equation}
  S_{\mathrm{pair}} = C_{\mathrm{seq}}\,\rho_{14},
  \qquad
  |\rho_{14}| = \frac{|S_{\mathrm{pair}}|}{|C_{\mathrm{seq}}|},
  \label{eq:wit:cal}
\end{equation}
a nonzero pathway-corrected pair-sector signal implies nonzero
collective DQ pair coherence and therefore functions as an
MQC/squeezing witness for the SU(1,1) pair sector.

\begin{remark}[Classical versus quantum cross-correlation]
\label{rem:classicalnoise}
It is important to emphasize that levels~(i) and~(ii) alone do not
formally distinguish between quantum non-separability and classical
statistical correlation. A shared, purely classical fluctuating
environment --- such as macroscopic classical magnetic field
fluctuations acting simultaneously on both sub-pools --- can also
generate a nonzero cross-diffusion block \(D_{AB}\) and produce
nonzero pair-sector signal content. Therefore, while levels~(i) and~(ii)
successfully witness the geometric boundary mechanism and the onset of
inter-spin correlation, they do not mathematically preclude a classically
correlated separable state. Definitively ruling out classical correlated
noise requires the strict threshold violation provided by the
level~(iii) many-body witness.
\end{remark}

\subsection{The bipartite thermal obstruction}
\label{sec:wit:obstruction}

If one attempts to evaluate entanglement strictly within a reduced
two-spin model, one encounters a severe thermodynamic obstruction. In a
high-temperature bulk NMR ensemble (such as living tissue at
\(310\,\mathrm{K}\)), the density matrix is heavily dominated by the
identity. The anti-aligned populations are approximately uniformly
distributed,
\[
  p_{\uparrow\downarrow}\approx p_{\downarrow\uparrow}\approx \tfrac14,
\]
so the separable boundary for two-qubit concurrence~\cite{Wootters1998}
is of order \(q\approx 0.25\).

However, macroscopic NMR signals are calibrated against the thermal
equilibrium magnetization \(M_0\), which scales with the tiny thermal
polarization \(\epsilon \sim 10^{-5}\). Even if the pathway-corrected
pair signal appears macroscopic relative to equilibrium, the absolute
magnitude of any single-pair off-diagonal density-matrix element remains
of order \(\epsilon\). Thus
\[
  |\rho_{14}| - q \approx 10^{-5} - 0.25 \ll 0.
\]
This is the pseudopure-state obstruction of liquid-state
NMR~\cite{Braunstein1999}: it is structural, not a calibration
limitation. The correct entanglement-theoretic setting is therefore not
a strictly bipartite reduced-spin witness but a collective MQC
framework.

\subsection{Level (iii): Many-body entanglement via MQC}
\label{sec:wit:mqc}

The MQC framework~\cite{Gaerttner2018} resolves the bipartite thermal
obstruction by evaluating the detected DQ-sector intensity against the
maximum achievable by a fully separable thermal ensemble. The
corresponding formal macroscopic witness takes the form
\begin{equation}
  \boxed{
  W_{\mathrm{MQC}}
  \equiv
  \frac{I_{\mathrm{DQ}}}{I_{\mathrm{sep}}^{\mathrm{bound}}}
  > 1
  \quad\Longrightarrow\quad
  \text{many-body SU(1,1) entanglement.}
  }
  \label{eq:wit:wmqc}
\end{equation}
Here
\[
  I_{\mathrm{DQ}} \propto \frac{|S_{\mathrm{pair}}|}{|C_{\mathrm{seq}}|}
\]
is the inferred DQ-sector MQC intensity and
\(I_{\mathrm{sep}}^{\mathrm{bound}}\) is the separable bound for the
same macroscopic thermal ensemble.

It is important to state why adapting the standard MQC entanglement
framework~\cite{Gaerttner2018} to the present non-compact setting is
theoretically nontrivial. Standard MQC bounds rely on the quantum
Fisher information and variance limits derived for compact SU(2)
ensembles, where the strictly bounded spectra of the angular momentum
operators guarantee finite fluctuation limits for any completely
separable state. By contrast, the SU(1,1) pair generators \(K_\pm\) are
non-compact and possess unbounded spectra. Deriving a rigorous,
universal separable bound \(I_{\mathrm{sep}}^{\mathrm{bound}}\) for
SU(1,1) therefore requires imposing explicit thermodynamic or
finite-energy cutoffs to mathematically constrain the maximal classical
variance. The formal derivation of this unbounded MQC threshold remains
an open problem in quantum metrology.

The detected signal is therefore interpreted first as a metric-regime
witness, second as an MQC/squeezing witness, and only under stronger
quantitative conditions as a many-body entanglement witness.

\subsection{Numerical status}

For the data of Ref.~\cite{KerskensPerez2022}, the measured signal
amplitude is of order
\[
  |S_{\mathrm{pair}}|_{\mathrm{meas}} \approx 0.15\,M_0.
\]
A further correction arises from transverse dephasing during the readout
interval. If the detected component decays with
\[
  T_2^* \approx 45\,\mathrm{ms},
  \qquad
  \tau \approx 45\,\mathrm{ms},
\]
then the measured amplitude is suppressed by approximately
\(e^{-1}\approx 0.37\), giving the decay-corrected estimate
\begin{equation}
  |S_{\mathrm{pair}}|_{\mathrm{corr}}
  \approx
  e^{\tau/T_2^*}|S_{\mathrm{pair}}|_{\mathrm{meas}}
  \approx
  0.41\,M_0.
  \label{eq:wit:T2corr}
\end{equation}

A definitive quantitative evaluation of \Cref{eq:wit:wmqc} ultimately
requires a full numerical simulation of \(C_{\mathrm{seq}}\) that
incorporates sequence imperfections such as finite pulse durations,
\(B_1\)-inhomogeneity, and relaxation during the gradient filter.
However, for the purpose of establishing a conservative lower bound on
the underlying pair-sector coherence, an exact experimental calibration
of \(C_{\mathrm{seq}}\) is not strictly necessary.

The pathway analysis of \Cref{sec:pathway} yields an idealized,
strictly unitary upper bound on the transfer efficiency,
\[
  |C_{\mathrm{seq}}^{\mathrm{ideal}}|
  \sim \frac18\,M_0.
\]
Any physical imperfection in the readout block can only reduce the
actual transfer efficiency,
\[
  |C_{\mathrm{seq}}^{\mathrm{actual}}|
  \le
  |C_{\mathrm{seq}}^{\mathrm{ideal}}|.
\]
Since the inferred pair-sector coherence scales inversely with this
coefficient,
\[
  |\rho_{14}| \propto |C_{\mathrm{seq}}|^{-1},
\]
using \( |C_{\mathrm{seq}}^{\mathrm{ideal}}| \) yields a strictly
conservative lower bound on the coherence required to account for the
measured signal.

With the decay-corrected amplitude
\[
  |S_{\mathrm{pair}}|_{\mathrm{corr}} \approx 0.41\,M_0,
\]
the inferred pair-sector coherence obtained from the idealized maximum
transfer efficiency is already macroscopically large on the NMR scale.
Any realistic sequence loss would only reduce
\(|C_{\mathrm{seq}}|\) further and therefore increase the required
underlying DQ-sector coherence.

This observation does not by itself close the many-body witness
analysis, because \Cref{eq:wit:wmqc} still requires an explicit
evaluation of the separable MQC bound
\(I_{\mathrm{sep}}^{\mathrm{bound}}\) for the present non-compact
SU(1,1) setting. What it does establish is that uncertainty in the
sequence-transfer calibration acts conservatively: improved calibration
can reduce the inferred coherence only up to the idealized upper-bound
transfer, whereas realistic imperfections can only strengthen the case
for a large underlying pair-sector MQC intensity.

Accordingly, if a future derivation of
\(I_{\mathrm{sep}}^{\mathrm{bound}}\) places the separable threshold
below this conservative lower-bound intensity, the resulting witness
violation will be robust against sequence-loss corrections.

%======================================================================
\section{Collective vs.\ Two-Subsystem Realisation}
\label{sec:bipartite}
%======================================================================

Whether the detected SU(1,1) nonclassicality constitutes entanglement
between distinguishable subsystems depends on how the generators are
physically realised.

For a single-mode or collective embedding,
\[
  K_+ = \tfrac12 a^{\dagger 2}, \qquad K_- = \tfrac12 a^2,
\]
violation signals internal squeezing within one collective Hilbert
space. For a two-mode realisation,
\[
  K_+ = a^\dagger b^\dagger, \qquad K_- = ab,
\]
the same class of witnesses probes correlations between separate
subsystems~\cite{HilleryZubairy2006,NhaKim2006}.

In the present setting,
\[
  K_\pm = I_{1\pm}I_{2\pm}
\]
generates a two-subsystem realisation. Several features support that
assignment:
\begin{enumerate}[label=(\alph*)]
  \item magic-angle suppression identifies anisotropic coupling between
  spatially distinct spin groups;
  \item linear scaling with \(M_0\) is consistent with selective pair
  coupling rather than collective single-manifold squeezing;
  \item refocusing at \(\Omega_+\) is characteristic of a pair mode
  linking distinguishable components;
  \item calibration of \(K_0 = \tfrac12(I_{1z}+I_{2z})\) resolves
  additive contributions from two distinguishable components.
\end{enumerate}

Under the two-subsystem assignment, violation of the MQC witness
\Cref{eq:wit:wmqc} would diagnose entanglement between two collective
spin populations rather than merely nonclassicality within a single
collective mode.

\subsection{Physical identity of the coupled subsystems}
\label{sec:bipartite:physics}

This subsection is best read as a secondary constraint analysis on
possible physical realizations of the two spin populations, not as the
primary theoretical content of the paper, which remains
covariance-geometric.

The algebraic framework treats the subsystems \(A\) and \(B\) abstractly as
two spin populations coupled to a common substrate. Their physical
assignment in brain tissue is not uniquely determined by the present
analysis, but it is heavily constrained by two independent experimental
observations: the suppression of the signal at the magic angle, and the
strict absence of conventional magnetization transfer (MT)
effects~\cite{KerskensPerez2022}.

\paragraph{What the subsystems cannot be.}

The magic-angle suppression requires an anisotropic interaction: the
underlying coupling must be dominated by residual dipolar couplings
whose angular dependence follows the \((3\cos^2\theta - 1)\) factor. In
purely isotropic bulk water, rapid molecular tumbling averages these
couplings to zero. The participating spins must therefore reside in a
structured, orientational restricted environment or a orientational gradient that can impose a asymmetry.

Separately, the absence of MT rules out the standard two-pool model of
clinical MRI, in which mobile water protons exchange magnetization with
the dense, restricted macromolecular proton pool (e.g., myelin sheaths,
protein backbones). If the pair coherence were formed through
interactions with this macromolecular pool, the signal would necessarily
exhibit MT-mediated cross-relaxation signatures. The empirical absence
of MT demonstrates that the SU(1,1) pair-sector signal does not arise
from classical thermodynamic exchange with the solid biological matrix.

These two eliminations are individually strong and jointly constraining:
isotropic bulk water is strongly disfavored by the magic-angle
dependence, and standard macromolecular-pool coupling is strongly
disfavored by the absence of MT.

\paragraph{The surviving constraint class.}

The subsystem assignment must therefore satisfy two simultaneous
conditions: the spin populations must experience residual dipolar
coupling (anisotropic, not motionally averaged), and they must not
participate in rapid proton exchange with the macromolecular backbone
(no MT). The remaining plausible constraint class therefore consists of
geometrically structured fluid domains that support residual dipolar
coupling without strong macromolecular exchange.

Candidate realisations within this class include kinetically isolated
but orientationally ordered water populations, such as specific layers
of interfacial hydration water, or water confined in anisotropic
nanostructures (e.g., highly ordered intra-axonal or interstitial
channels). In such environments, water molecules exhibit residual
dipolar couplings --- satisfying the magic-angle constraint --- but
lack the rapid proton exchange with the macromolecular backbone that
would drive bulk MT.

The specific identification of the two populations within this
constraint class remains an open experimental question. Future
experiments could narrow the assignment further: for instance,
diffusion-weighted variants of the pulse sequence could test whether the
pair signal depends on restricted diffusion geometry, which would
discriminate between candidate realisations by their characteristic
diffusion length scales.

\paragraph{Architecture of the shared-environment coupling.}

Regardless of which specific candidate is realised, the constraint
analysis clarifies the physical architecture of the transfer mechanism.
The ``shared environment'' that acts as the physical conduit for the
boundary-conditioned transfer (\Cref{sec:bridge:coupling}) is the
dynamic biological matrix --- the tissue substrate whose covariance
evolves on the Bures--Wasserstein manifold. The two spin populations
that receive the transferred cross-mode structure, however, are
fluid-based: they are water-proton domains embedded in, but
kinetically isolated from, the macromolecular scaffold.

This architecture also provides an additional discrimination against
classical cross-relaxation artifacts. If the pair signal were driven by
conventional MT or chemical-exchange pathways, the absence of MT
signatures would be inexplicable. The fact that the signal coexists
with strictly conventional MT behavior supports the interpretation that
the pair coherence arises from boundary-conditioned geometric transfer
through the shared substrate environment rather than from classical
thermodynamic exchange with the macromolecular pool.

%======================================================================
\section{Discussion}
\label{sec:discussion}
%======================================================================

This paper completes the measurement-theoretic arm of a three-paper
programme. Paper~1 develops the Bures--Wasserstein geometry of Gaussian
covariance matrices and proves the constructive noise bound that governs
cross-mode structure at the Williamson floor~\cite{Paper1}. The
companion paper shows that substrate-constrained covariance flow in the
living human brain can reach that boundary regime, where single-mode
compression saturates and further admissible covariance evolution is
forced into the cross-mode complement~\cite{MetricPaper}. The present
paper derives how that boundary transition becomes experimentally
visible in an embedded spin probe and shows that the reported magnetic
resonance data exhibit signatures consistent with the resulting
mechanism.

The central theoretical result is the boundary-conditioned transfer
theorem (\Cref{prop:inherit}). The key claim is not that the reduced
spin system itself reaches the same Williamson-boundary point as the
substrate. Rather, once the substrate reaches the boundary, the
admissible continuation of its covariance flow becomes cross-mode in the
saturating sector, and it is this boundary-conditioned cross-mode
structure that is transferred into the reduced spin dynamics. In the
spin probe, the transferred structure appears as a nonzero cross-mode
block \(G_{AB}^{\mathrm{spin}}\), with leading-order action in the
inter-spin correlation sector rather than in the ordinary marginal
relaxation sector.

The transfer mechanism has been formulated in a weak-coupling,
shared-environment, Born--Markov reduction, together with a local
linear-Gaussian approximation for the boundary-conditioned correction to
the diffusion tensor. Within that approximation class, the shared
substrate environment provides the physical map from substrate
cross-mode covariance to reduced spin cross-diffusion. The paper
therefore does not present a model-independent microscopic derivation of
the full reduced generator; rather, it identifies a concrete and
internally consistent open-system regime in which the geometric boundary
structure of the substrate becomes visible to the spin probe through the
correlated sector of the reduced dynamics.

The algebraic analysis then shows how this transferred spin cross-mode
structure decomposes inside the two-spin operator algebra. The relevant
recipient sectors are the compact SU(2) exchange manifold in the
zero-quantum sector and the non-compact SU(1,1) pair manifold in the
double-quantum sector. The projection formulas derived in
\Cref{sec:algebra:projection} make this decomposition explicit, while
the symmetry-selection argument of \Cref{sec:algebra:symmetry} explains
why boundary-forced squeezing is expected to populate the DQ/SU(1,1)
pair sector preferentially. That selection argument remains
conditional on the transferred substrate cross-mode block being
predominantly of squeeze type and on the transfer preserving the
relevant quadrature-parity structure to leading order. Under those
conditions, however, the observed refocusing at the pair frequency
\(\Omega_+\), rather than the exchange frequency \(\Omega_-\), strongly
supports the predicted sector selection.

The pulse-sequence analysis then resolves the experimental detection
problem. Because the dominant algebraic recipient is the DQ/SU(1,1)
pair sector, the induced coherence is not directly preserved by the
gradient-filtered readout. Instead, the
\(45^\circ\)--gradient--\(45^\circ\) block converts a portion of the DQ
pair coherence into a gradient-surviving \(p=0\) intermediate and then
into detectable single-quantum magnetization. The crusher gradient also
acts as a directional spatial selector, so that the detected pathway is
sensitive not only to coherence order but also to the spatial geometry
of the dipolar pair sector. In this framework, the observed
magic-angle suppression is naturally interpreted as evidence that the
detected signal depends on anisotropic dipolar-mediated pair structure
rather than on an orientation-insensitive artifact.

Taken together, these results establish a continuous chain from geometry
to experiment:
\begin{center}
  substrate BW boundary \(\to\)
  boundary-conditioned cross-mode continuation \(\to\) \\
  transferred spin cross-mode structure \(\to\)
  DQ/SU(1,1) pair-sector projection \(\to\) \\
  readout-converted NMR signal.
\end{center}
The distinctive claim of the paper is that the observed signal is best
understood as the spin-probe signature of this full chain, not as an
isolated pulse-sequence artifact and not as a conventional compact
exchange phenomenon.

Conceptually, the priority runs from covariance geometry to physical
embodiment, not the reverse: the physical mechanism matters here as the
mode of observability of a boundary-limited informational regime.

At the interpretive level, the paper supports a three-level hierarchy.
First, the signal functions as a metric-regime witness: it is
consistent with the substrate entering the deep Bures--Wasserstein
boundary regime in which single-mode compression is exhausted and
cross-mode continuation becomes necessary. Second, the signal functions
as a macroscopic MQC/squeezing witness: after pathway correction, it
indicates nonzero collective DQ coherence in the non-compact SU(1,1)
pair sector of the spin probe. These two levels constitute the
quantitatively established achievements of the present work, with the
caveat that levels~(i) and~(ii) do not by themselves formally
distinguish quantum non-separability from classical statistical
correlation (\Cref{rem:classicalnoise}).

The strongest level of the hierarchy --- promotion of the signal to a
many-body entanglement witness --- remains formal rather than numerically
closed. A key clarification of the present analysis is that a strictly
bipartite reduced two-spin witness is not the appropriate framework for
the room-temperature bulk-NMR setting. Because the density matrix is
identity-dominated, any isolated-pair entanglement criterion is
thermodynamically obstructed by the pseudopure-state problem
\cite{Braunstein1999}. The correct entanglement-theoretic setting is
therefore the macroscopic MQC framework, in which the detected DQ-sector
intensity must be evaluated against the separable bound for the full
thermal ensemble rather than for a single isolated pair.

Accordingly, the present paper should be read as providing the formal
entanglement-theoretic framework required for a future many-body
entanglement claim, rather than as already establishing that claim in
full. Definitive closure would require a full numerical simulation of
the transfer coefficient \(C_{\mathrm{seq}}\), computation of the
separable MQC bound \(I_{\mathrm{sep}}^{\mathrm{bound}}\) for the
non-compact SU(1,1) DQ sector, and sequence-variation experiments that
separate pair-sector and compact-exchange contributions through their
distinct refocusing behavior. These are concrete calibration tasks, not
unresolved conceptual contradictions.

The conditional structure of the boundary-conditioned transfer theorem
provides clean falsifiability. If the companion paper predicts boundary
saturation under specific physiological conditions, then the spin probe
should exhibit the SU(1,1) pair signal. Absence of the signal would
falsify either the substrate model or the transfer mechanism.
Conversely, observation of the signal under conditions where the
substrate is not predicted to approach the boundary would indicate an
alternative mechanism. The cardiac locking and the internal structure of
the detected signal provide an additional discrimination principle:
heartbeat correlation alone is insufficient, but the conjunction of
heartbeat locking with pair-frequency refocusing, even-echo parity, and
the DQ-to-SQ readout pathway is substantially more difficult to
reconcile with straightforward classical pulsatile alternatives.

The physical identity of the coupled spin subsystems is constrained but
not uniquely determined by the present analysis
(\Cref{sec:bipartite:physics}). The combination of magic-angle
dependence and absence of MT signatures eliminates both isotropic bulk
water and the standard macromolecular two-pool model, constraining the
subsystems to geometrically structured fluid domains with residual
dipolar coupling but without macromolecular exchange. Future
diffusion-weighted variants of the pulse sequence could narrow this
assignment further.

If confirmed, the mechanism developed here would show that spin-probe
magnetic resonance can register not merely local relaxation parameters
but features of the covariance-geometric regime of the substrate on the
Bures--Wasserstein manifold. In that sense, the DQ/SU(1,1) pair-sector
signal would serve as a concrete spin-probe signature of the deep
boundary regime in the living human brain.

\appendix

%======================================================================
\section{Hyperbolic evolution and compact algebras}
\label{app:hyperbolicity}
%======================================================================

Under compact \(\mathfrak{su}(2)_{\mathrm{ZQ}}\) evolution generated by
\(H = J(I_{1+}I_{2-}+I_{1-}I_{2+})\), any expectation value
\(\langle O(t)\rangle\) with
\(O \in \mathfrak{su}(2)_{\mathrm{ZQ}}\) satisfies
\[
  \langle O(t)\rangle = \sum_k c_k e^{i\lambda_k t},
\]
where the adjoint eigenvalues \(\lambda_k\) are purely imaginary.
Hence the evolution is a finite sum of oscillatory terms; hyperbolic
growth is impossible within the compact algebra.

For \(\mathfrak{su}(1,1)\), the Killing form has indefinite signature,
the adjoint representation admits real eigenvalues, and the evolution
includes \(\cosh(gt)\) and \(\sinh(gt)\) terms.

%======================================================================
\section{Scaling of SU(1,1) signal with \texorpdfstring{$M_0$}{M0}}
\label{app:scaling}
%======================================================================

For a collective single-manifold SU(1,1) realisation with
\(K_+ = \tfrac12 a^{\dagger 2}\), the pair amplitude scales as
\(\langle K_+\rangle \propto N(N-1)\propto M_0^2\). For a bipartite
two-subsystem realisation with \(K_+ = a^\dagger b^\dagger\), the pair
amplitude scales as \(\propto M_0\). The approximately linear scaling
observed in the data therefore favors the bipartite interpretation.

%======================================================================
\section{Derivation of the cross-mode projection}
\label{app:projection}
%======================================================================

Using
\[
  I_{\alpha x} = \tfrac12(I_{\alpha+}+I_{\alpha-}),
  \qquad
  I_{\alpha y} = \tfrac{1}{2i}(I_{\alpha+}-I_{\alpha-}),
\]
together with
\[
  K_\pm = I_{1\pm}I_{2\pm},
  \qquad
  J_\pm = I_{1\pm}I_{2\mp},
\]
one obtains
\begin{align}
  I_{1x}I_{2x}
  &= \tfrac14(K_+ + K_- + J_+ + J_-)
   = \tfrac12 K_1 + \tfrac12 J_1,\\
  I_{1y}I_{2y}
  &= -\tfrac14(K_+ + K_- - J_+ - J_-)
   = -\tfrac12 K_1 + \tfrac12 J_1,\\
  I_{1x}I_{2y}
  &= \tfrac{1}{4i}(K_+ - K_- - J_+ + J_-)
   = \tfrac12 K_2 - \tfrac12 J_2,\\
  I_{1y}I_{2x}
  &= \tfrac{1}{4i}(K_+ - K_- + J_+ - J_-)
   = \tfrac12 K_2 + \tfrac12 J_2.
\end{align}
Collecting these expressions into matrix form yields
\Cref{eq:proj:DQZQ}.

\section*{Acknowledgments}

The author acknowledges the use of AI assistants for structural
brainstorming, language refinement, and \LaTeX\ drafting support. The
author bears full responsibility for the scientific content, arguments,
and equations presented herein.

\bibliographystyle{abbrv}
\bibliography{bib}

\end{document}